# Creating A Galactic Plane Atlas With Amazon Web Services

G. Bruce Berriman[1*], Ewa Deelman[2], John Good[1], Gideon Juve[2], Jamie Kinney[3], Ann Merrihew[3], and Mats Rynge[2]

[1]Infrared Processing and Analysis Center, California Institute of Technology, Pasadena, CA 91125, USA
[2]Information Sciences Institute, University of Southern California, Marina Del Rey, CA 90292, USA
[3]Amazon Web Services, 1918 8th Ave, Seattle, WA 98101, USA

Email: {gbb, jcg}@ipac.caltech.edu; {deelman, gideon, rynge}@isi.edu; {jkinney,amerrihew}@amazon.com

\* Corresponding author

## Abstract

This paper describes by example how astronomers can use cloud-computing resources offered by Amazon Web Services (AWS) to create new datasets at scale. We have created from existing surveys an atlas of the Galactic Plane at 16 wavelengths from 1 μm to 24 μm with pixels co-registered at spatial sampling of 1 arcsec. We explain how open source tools support management and operation of a virtual cluster on AWS platforms to process data at scale, and describe the technical issues that users will need to consider, such as optimization of resources, resource costs, and management of virtual machine instances.

## 1. Introduction: Creating The Galactic Plane Atlas

How can astronomers who are inexpert in technology take advantage of cloud computing technologies? And how can astronomers optimize performance and minimize costs under the pay-as-you-go tariffs for processing and storage offered by cloud computing providers? We have set out to answer these questions by using open source tools and methods to create and operate a virtual cluster on the Amazon Elastic Cloud 2 (hereafter, EC2) of Amazon Web Services (AWS) and create an imaging Atlas of the galactic plane at 16 infrared wavelengths from 1 μm to 24 μm.

The Atlas has been created from input images from five major surveys, listed in Table 1 and archived at the NASA/IPAC Infrared Science Archive (IRSA), processed with the Montage image mosaic engine [1] (see also http://montage.ipac.caltech.edu), a toolkit that performs all the steps to create mosaics for a set of input files covering an area of the sky. Pixel data in all images have been transformed to 1 arcsec spatial sampling in the Cartesian projection, co-registered on the sky, and represented in Galactic coordinates. The Atlas thus appears to have been measured with a single instrument and telescope operating across the full wavelength range, incorporating all those data within Galactic coordinates $l=\pm360°$ and $b=\pm20°$; Table 1 includes the coverage of each survey within this area. The Atlas is organized and stored in tiles 5° x 5°, with tile centers separated by 4°. The final output data set size has a volume of 45 TB. The organization provides a 1° overlap between tiles aimed at supporting validation. Following completion of validation, the Atlas will be archived at AWS, and made publicly accessible in spring 2014 through an Applications Programming Interface (API).

**Table 1: The surveys and bands included in the Galactic Plane Atlas, including output data sizes and compute times on the Amazon Cloud.**

| Survey | Bands (μm) | Coverage of Atlas area (%) | Output Size (TB) | Compute Time [1] (1,000's core hours) |
|---|---|---|---|---|
| **2MASS** | 1.2, 1.6, 2.2 | 100 | 14.4 | 87 |
| **GLIMPSE** | 3.6, 4.5, 5.8, 8.0 | 11 | 2.0 | 60 |
| **MIPSGAL** | 24 | 8 | 0.4 | 3 |
| **MSX** | 8.8, 12.1, 14.6, 21.3 | 35 | 6.8 | 36 |
| **WISE** | 3.4, 4.6, 12, 22 | 100 | 19.2 | 132 |

[1] On the EC2 hi1.4xlarge EC2 instance. See text for details.

EC2 uses virtualization technologies to offer essentially unlimited on-demand, pay-as-you-go access to compute and storage resources, which are released on completion. Astronomers are generally not skilled at managing and optimizing the environments of clusters of virtual machines, and we describe here how well open source tools can successfully manage many of these tasks on behalf of the user. Consider throughout that in the pay-as-you-go cost model, operating inefficiencies may well lead to substantially increased costs.

## 2. Managing Workflows For The Galactic Plane Atlas

Creation of the galactic plane atlas is an example of a highly parallelizable, data driven workflow, where the output of one step becomes the input to the next, and where each step can be parallelized across as many machines as are available. Executing such a workflow on a distributed platform is feasible but tedious [3], for it involves, among things, locating compute and storage resources, scheduling jobs and managing failures

Our recommended approach is to take advantage of workflow management systems (hereafter, WMS), designed to perform these tasks. These tools take descriptions of workflows and set up and run a processing plan for an execution environment whose configuration and organization is incorporated into the workflow manager. In this way, workflow managers abstract the details of the operation of a virtual cluster away from the user. We have used in this study the Pegasus WMS [4] (see also http://pegasus.isi.edu), a mature, highly-fault tolerant system used in many disciplines.

A description of the workflow – that is, the data flow and processing paths (including dependencies between steps)– are represented in XML format as a Directed Acyclical Graph (DAG) and are created by APIs included in Pegasus. Pegasus takes this DAG and creates an executable workflow optimized for the target environment, in this case EC2. Figure 1 shows the organization of one of 16 hierarchical workflows, one for each wavelength.

Pegasus is layered on top of two open source tools required to run the workflows. HTCondor (http://research.cs.wisc.edu/htcondor/index.html) performs job queuing and scheduling, and DAGMan (Directed Acyclic Graph Manager; http://research.cs.wisc.edu/htcondor/dagman/dagman.html) ensures that jobs are submitted to HTCondor in the order implied by the dependencies encoded in the DAG.

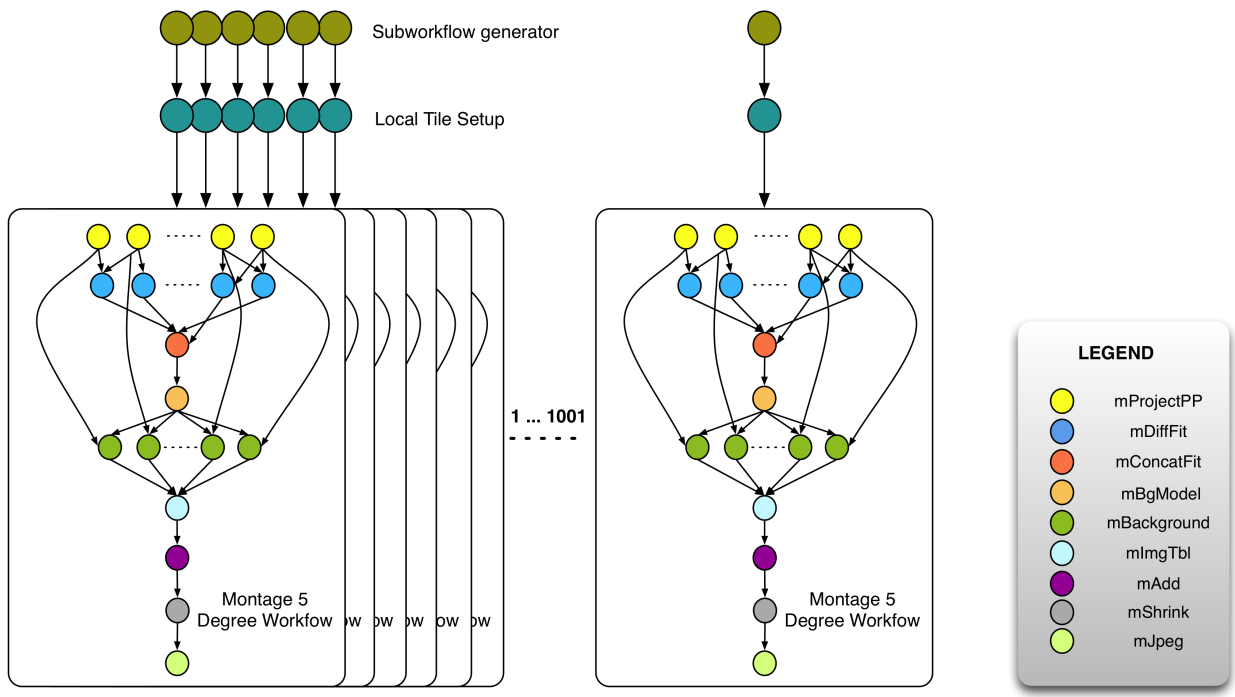

Fig 1: A representation of the creation of one of the 16 workflows used to create the Galactic Plane Atlas. The colored circles represent one step in the processing, with the Montage components named in the legend. Each workflow contains 1,001 sub-workflows, one for each 5 x 5 tile.

## 3. Running the Workflows on the Amazon Web Services Cloud.

Given that Montage and Pegasus are designed to run on all common Unix-based platforms, our technical approach has been to create an environment within a virtual machine in which these applications run, and subsequently replicate this machine across nodes of a cluster. (an alternative approach is to adapt applications to a MapReduce framework such as Hadoop; e.g. [2]). Creating and configuring virtual machine images and organizing them into a distributed cluster is, however, tedious and requires system administration knowledge.

The tool set described above – Montage, Pegasus, and HTCondor (along with dependent services) – was copied and built on a virtual machine image that is loaded on to the EC2 virtual machine, which was then configured for the correct networking and system setup. Unless users have systems management experience, we strongly recommend they take advantage of existing images – one for EC2 is provided on the Pegasus web page – or use a service such as Puppet (http://puppetlabs.com/) or Chef (http://docs.opscode.com/index.html) to construct and manage them. Creating a virtual cluster manually, by launching and configuring individual nodes is also not recommended. Tools such as Wrangler (http://www.isi.edu/~gideon/publications/JuveG-Automating.pdf), can easily automate the provisioning and configuration of clusters running on Amazon EC2.

Figure 2 shows the architecture of the virtual cluster and storage created for the Galactic Plane Atlas. HTCondor requires the master/worker architecture shown there. The master instance is used for submitting workflows to the worker nodes for processing. The worker nodes used Amazon Simple Storage Service S3 (hereafter, S3) object storage, supported by Pegasus, to provide a centralized storage system to house input data, temporary storage for intermediate data, and secure place for the final data products. S3 is a highly efficient choice for storage as it scales with the number of clients, at the cost of some latency in each request. Ephemeral disks were used in a striped RAID-0 configuration to increase the local disk I/O performance.

Processing performance was limited by the data transfer rate from IPAC. To manage data transfer rates, we used a reverse Squid caching server (http://www.squid-cache.org/) installed at ISI, as indicated in the bottom frame of Figure 3. We added and removed nodes manually as needed in response to the availability of IPAC network egress bandwidth, rather than allow the cluster to change size in response to processing demand. This provisioning strategy maintains the system at a size that the data rates can sustain. Compute nodes achieved close to 100% CPU utilization by oversubscribing the worker nodes with job. The over subscription was done in order to better overlap the CPU intensive computations and manage the somewhat high latency in data transfers to and from S3.

Thus, while Pegasus provided a robust framework for creating and executing the workflows, fully optimizing the processing required human intervention and knowledge of how to design virtual machines and how the workflow management tools.

## 4. Processing Costs

Users should ideally perform a cost benefit analysis or benchmarking to understand which EC2 instances will meet their performance goals and budget constraints. We used the EC2 *hi1.4xlarge* instance type, and this consumed 318,000 core hours to complete the entire set of workflows over all 16 bands, and this would have cost $5,950 using spot pricing (essentially, bidding on unused capacity).  Comparison tests show that if we had chosen to run the workload on the *cc2.8xlarge* instances with spot pricing, it would reduce the total cost to $2,200 based on the lower cost per core/hour and faster processor speed. These costs do not include storing the resulting data set in Amazon S3.

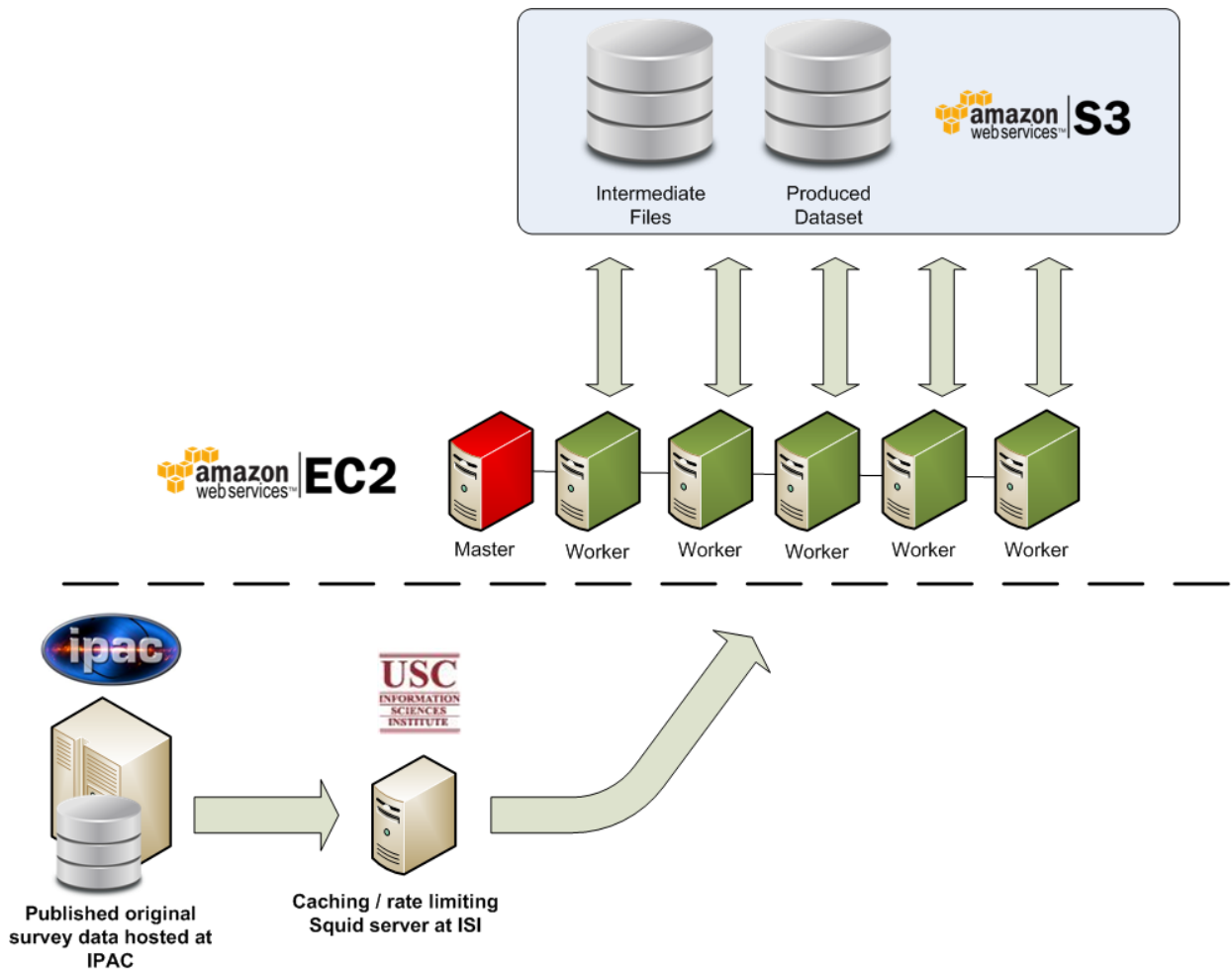

Fig 2: The architecture of the processing environment. In the Amazon EC2 cloud,
A master instance is used to start workflows on worker nodes, which use centralized Amazon S3 storage. Outside the cloud, a squid server installed at ISI controls the rate of data transfer from IPAC to EC2.

## 5. Conclusions

The creation of the Galactic Plane Atlas shows that Amazon EC2 is a powerful resource for astronomical computing at scale, and open source tools are invaluable for taking advantage of it. Nevertheless, managing a virtual environment, optimizing processing and controlling costs requires investment in understanding how AWS, Linux systems and the workflow tools operate. The pay-as-you – go cost model requires cost benefit analyses to understand the cost and performance of AWS' processing and storage options.

## Acknowledgements

Compute resources for the project were provided by an AWS in Education Research Grant (http://aws.amazon.com/grants/). Storage for the Galactic Plane data set is provided by the AWS Public Data Sets program (http://aws.amazon.com/datasets/). This paper is based upon work supported in part by the National Science Foundation under Grants (OCI-0943725, 0910812). G. B. Berriman and J. C. Good are supported by the NASA Exoplanet Science Institute at the Infrared Processing and Analysis Center, operated by the California Institute of Technology in coordination with the Jet Propulsion Laboratory (JPL).

Bruce Berriman is a Senior Scientist at IPAC, Caltech (PhD in Astronomy, Caltech in 1983). His research centers on investigating the applicability of emerging technologies to astronomy.

Ewa Deelman is a Project Leader at USC/ISI (PhD in Computer Science, RPI in 1997). Her research interests include distributed scientific environments, with emphasis on workflow management.

John Good is a distributed science information systems architect at IPAC (PhD in Astronomy, U. Mass, 1983). His focus is on scalable architectures for handling extremely large datasets, especially astronomical imaging and relational catalogs and on science visualization/user interfaces.

Gideon Juve is a Computer Scientist at USC/ISI (PhD in Computer Science, USC in 2012). His research focuses on enabling and optimizing scientific workflows on clusters, grids and clouds.

Jamie Kinney leads the HPC and Scientific Computing team at Amazon Web Services (BS in Marine Science/Biology, University of Miami in 1996). His work focuses on the application of cloud technologies to a broad range of scientific workloads

Ann Merrihew is an Amazon Web Services (AWS) Account Manager for Higher Education in California, and focused on technology and Education previously at Microsoft (2004-2011) and Apple (2011-2012) prior to joining AWS in early 2012.

Mats Rynge is a Computer Scientist at USC/ISI (BS in Computer Science, UCLA in 2002). His research interests include distributed and high performance computing.